\documentclass[12pt]{elsarticle}

\usepackage{amssymb}
\usepackage{graphicx}

\newcommand{\app}{\setcounter{section}{0}
\setcounter{equation}{0} \renewcommand{\thesection}{Appendix
\Alph{section}}\renewcommand{\theequation}{\Alph{section}.
\arabic{equation}}}
\journal{Physica A}

\begin{document}

\begin{frontmatter}
\title{Non-extensive thermostatistics approach to metal melting entropy}

\author{P. Quarati$^1$ and A.M. Scarfone$^2$}

\address{$^1$ Dipartimento di Scienze Applicate e Tecnologia, Politecnico di
Torino, Corso Duca degli Abruzzi 24, I-10129, Italy and Istituto
Nazionale di Fisica Nucleare, Sezione di Cagliari, I-09042 Monserrato, Italy\\
$^2$ Istituto dei Sistemi Complessi - Consiglio Nazionale delle Ricerche (ISC-CNR) c/o Politecnico di Torino, Corso Duca degli Abruzzi 24, I-10129, Italy}
\date\today

\begin{abstract}
A generalized Mott relation of metal melting entropy is derived by means of non-extensive solid and liquid quantum entropy that we calculate from grand partition functions of localized ordered quantum solid and of disordered quantum Boltzmann liquid. For each of the 18 elements considered the entropic parameter $q_m$, depending on particle correlations, is deduced such that a better agreement is obtained between calculated non-extensive metal melting entropy and available experimental data. The non-extensive entropic parameter makes the difference between normal and anomalous metals. Therefore, also those not reported here should belong to one of the two classes. Possible applications to condensed matter, Earth and other solar planets seismology are mentioned.
\end{abstract}

\begin{keyword}
Quantum statistical mechanics; Solid-liquid transitions; Melting of specific substances
\end{keyword}

\end{frontmatter}

\section{Introduction}
The melting process of metals, semiconductors, alloys and nanocrystals
is investigated for its intrinsic importance and to understand thermodynamics of phase transitions, condensed matter structures, planets geophysical structure, geology, astrophysics, dynamics of solar planets and inner Earth core.\\
Well known are the semi-empirical rules introduced by Sutherland \cite{Sutherland} and Lindemann \cite{Lindemann} and their amendments \cite{Wallace,Regel}, validated by the application to melting of the atomic Einstein-Mott model to evaluate, at melting temperature, atomic vibration amplitude relative to atomic spacing.
Melting takes place when the atomic oscillation amplitude is larger than the interatomic distance, the change in the nature of the vibration spectrum is the only reason for melting \cite{Mott,Lawson}.

After quick progress in early times, when many authors understood the
mechanism of metal melting \cite{Frenkel,Faber}, more recently the process
was described by Regel' and Glazov \cite{Regel}, among others. They showed that the vibrational entropy change is the main contribution
to metal melting entropy, while for semiconductors the electronic structure change from solid to liquid is also a leading contribution. Wallace
\cite{Wallace} described the melting process introducing all the effective liquid correlation terms and observed that metals can be divided
into two groups: normal when the electronic ground state between solid and liquid phases does not change, anomalous when the electronic ground state changes and the liquid higher order correlation contributions are not negligible.\\
The metal melting entropy can be given in a simple form as the logarithm of solid and liquid frequencies related to the Einstein-Mott vibration atomic model, allowing to link melting entropy and electrical conductivity. Other models have been developed in \cite{Dash}. Theoretical studies to understand how the gradual process of melting evolves, a problem studied also by Feynmann as reported by Dash \cite{Dash}, consolidate the microscopic formulation of the melting entropy as a logarithmic function.

In the solid-liquid transition the order due to atomic localization in the solid phase is lost. At melting, the solid phase is metastable and, in the liquid, disordered phase correlations are active. Therefore, models of melting entropy require the calculation of many different contributions. In these frames, our thought is that metal melting entropy more conveniently can be
described by means of the generalized non-extensive (NEXT) statistical mechanics as developed in recent years by \cite{Tsallis,Tsallis1,Tsallis2,Tsallis3} and by others authors \cite{CEJP,MPLB,PTP}. In their approaches, an entropic parameter takes into account the presence of correlations in quasi-stationary or metastable systems. This parameter can make the difference among different values of metal melting entropy.\\ After the calculation of the solid and liquid NEXT entropy $S_{2-q}$ through the respective grand partition functions using the
approaches of \cite{Fowler2}, we have evaluated the melting
quantum entropy of metals, modified with the recipes of the NEXT statistical mechanics, as a function of the solid and liquid atomic frequencies
derived from the measured conductivity \cite{Regel,Fowler2,Hultgren,Chi}. It is significant, for the validity of NEXT statistical mechanics in quasi-stationary states with correlations, to verify that a good agreement with experimental results of melting entropy can be reached by the simple NEXT modified Mott relation, avoiding calculation of correlation contributions.\\
Let us outline the aims of this work. In the melting of a metal (and vice versa in freezing) we have a quantum localized system, described by a localized distribution, that, at melting temperature, becomes delocalized and is described by a quantum Boltzmann distribution \cite{Fowler2}. During melting, a solid system (crystal), with almost no correlation in a metastable state, is transformed to a system (liquid) with many local correlations (vice versa during freezing).\\
At melting, the two states (quantum localized and quantum Boltzmann) coexist with the same entropic parameter $q_m$, at temperature $T_m$. We can consider that temperature could fluctuate around its average value. In fact, the liquid system has a different volume than the solid phase. Variation of volume from the metastable solid state to liquid is connected with particle correlations i.e. with density fluctuation and then with temperature fluctuation.
Furthermore, particle correlations are responsible of density fluctuations and therefore of temperature fluctuations that gives rise to non-extensivity. In this sense, particle correlations are equivalent to temperature fluctuations. In fact, as shown in \cite{Corraddu}
the parameter $q$ is strictly related to particle correlations according to
$q=1\pm 24\,\Gamma^2\,\alpha^4$ that arises in the study of thermonuclear reactions in the solar core plasma, where $\Gamma$ is the plasma parameter and $\alpha$ a nucleon-nucleon correlation parameter.
Therefore, the consequence is that the non-extensive treatment is fully justified \cite{Tsallis}.\\
After comparing the NEXT evaluation of metal melting entropy with the experimental results we can analyze the behavior of the different metals with respect to the entropic parameter $q_m$. We show that metals can be grouped in two families with two different values of $q_m$.
The entropic parameter makes a difference between correlations present in normal and anomalous metals. The possible division of metallic elements into two groups was already indicated in \cite{Wallace,Regel}.

By considering that: i) during the transition from the solid phase to the liquid one the metal goes from an ordered, localized solid to a disordered liquid system; ii) the process is cooperative and in the liquid there are many-body correlations and also memory effects in intermediate state seem to appear; iii) melting is a first order transition that is not exactly at equilibrium, or at a thermodynamic global equilibrium, but rather at a meta-equilibrium or quasi-stationary state; iv) the melting entropy is related to the logarithmic function of the grand canonical potential, it is natural to advance the hypothesis that metal entropy could more conveniently be described as a $q$-logarithmic function ($\ln_q x$) of the grand canonical potential within the frame of NEXT statistical mechanics whose use in this context we want to justify and validate.\\
Finally, we remark that Boltzmann entropy does not take into account particle correlations while deformed entropy does through the deformation entropic parameter and shows as this parameter can distinguish between normal and anomalous metals.\\
Let us conclude this paragraph by summarizing the content of this work. We first introduce the classical Mott relation, then we derive melting
quantum entropy by means of quantum partition functions of localized and
quantum Boltzmann distributions. Then we present the NEXT modified Mott relation. We discuss the calculated and the observed results and give our
interpretation of them. Because the correction to Mott melting entropy depends on the average number $N$ of atoms of the ensemble we report in a figure the dependence of metal melting entropy at given $q_m$ on $N$. Finally, we conclude recalling the importance of metal melting also in planet
cores and in some astrophysical objects as well as in nanostructures,
semiconductors and superconductors.\\

\section{Classical approach}
In standard statistical mechanics, based on
Boltzmann-Gibbs (BG) approach (extensive thermodynamics), melting proceeds at constant
$T_m$ and can be studied simply in the grand canonical formalism by equalling solid and liquid Gibbs free
energy.

We start by introducing the extensive entropy according to
\begin{eqnarray}
  S^{\rm BG}=-k_{\rm B}\sum_{i=1}^Wp_i\,\ln p_i \ ,\label{SBG}
\end{eqnarray}
where $W$ is the number of possible configurations occurring with probability $p_i$ and $k_B$ is the Boltzmann constant.\\
Sometime, especially in chemistry texts, Eq. (\ref{SBG}) is named configurational or positional entropy. In the microcanonical picture where all different configurations are realized with the same probability $p_i=1/W$, Eq. (\ref{SBG}) measures the number of possible configurations of the system: $S^{\rm BG}=k_{\rm B}\,\ln W$. Differently, thermodynamic entropy is defined through the relation $dS=\delta Q/T$ where $\delta Q$ is the infinitesimal heat exchanged by the system with its environment at temperature $T$. However, it is strongly believed that configurational entropy and thermodynamical entropy should represent ultimately the same physical quantity in spite of their different expression \cite{Lambert}.\\
Differently, in the non-extensive formalism, the equivalence between the configurational entropy and thermodynamical entropy is still an open question. Notwithstanding, under the condition of {\em no work}, a situation encountered in the study of melting, the fundamental thermodynamical relation of $T\,dS=\delta Q$ holds also in the non-extensive picture, justifying reasonably the equivalence between the configurational and thermodynamical entropy also in the NEXT formalism. Consequently, in the following we assume that vibrational entropy equal to the measured thermal entropy.\\
In the grand canonical picture, equilibrium distribution is obtainable by maximizing entropy (\ref{SBG}) for a fixed value of the mean energy $U$ and the mean particle numbers $N$. It turns out to be
\begin{eqnarray}
p_{j,n}={1\over\Xi}\exp\left(-{\epsilon_j-\mu n\over k_{\rm B}T}\right) \ ,
\end{eqnarray}
for a given energy level $\epsilon_j$ and particles number $n$, where $\mu$ is the chemical potential.\\ In this way, BG entropy at equilibrium assumes the expression
\begin{equation}
  S^{\rm BG}=k_B\left(\ln\Xi+{U\over k_{\rm B}T}-N\ln z\right) \ ,\label{S}
\end{equation}
whereas $U=\sum_{j,n}\epsilon_j p_{j,n}$ and $N=\sum_{j,n}n p_{j,n}$ are related to the partition function by
\begin{eqnarray}
&&N=z{\partial\ln\Xi\over\partial z} \ ,\\ &&U=k_BT^2{\partial\ln\Xi\over\partial T} \ ,
\end{eqnarray}
and $z=e^{\mu/k_{\rm B}T}$ is the fugacity.

To study melting, two different grand canonical partition functions are used to describe the solid and
liquid phases. They are given by
\begin{eqnarray}
\Xi_S(z,\,T)=\left(1-zf_S(T)\right)^{-1} \ ,
\end{eqnarray}
with $zf_S(T)<1$, and
\begin{eqnarray}
\Xi_L(z,\,T)=\exp(zf_L(T)) \ ,
\end{eqnarray}
for the solid and liquid phases, respectively, where $f_S(T)$ and $f_L(T)$ are the corresponding partition functions of single particle.\\ The solid phase, can be depicted by the Einstein model for a mono-atomic metal with atoms, like harmonic oscillators, localized in the $N$ nodal positions of the lattice crystal. In this situation the single particle canonical distribution reads
\begin{eqnarray}
f_S(T)=\left({k_BT\over h\nu_S}\right)^3\exp\left(-{\epsilon_S\over k_BT}\right)
\ ,\label{distrs}
\end{eqnarray}
and describes the statistics of a quantum oscillator in the limit of high temperature
with $k_BT\gg 1/h\nu_S$, where $\nu_S$ is the vibration frequency of molecules, $\epsilon_S$ is the fundamental energy of each oscillator and $h$ the Planck constant.\\
Differently, in the liquid phase each atom is non-localized and can occupy any of the $N$ nodal position so that the related distribution function is $f_L(T)=Nf_S(T)$, with $\nu_L$ and $\epsilon_L$ being the corresponding quantities for the liquid phase.

Under the condition of no external work, the melting heat is given by the difference of internal energy of liquid and solid at the melting temperature $T_m$
\begin{eqnarray}
\Lambda=U_L-U_S\equiv N\left(\epsilon_L-\epsilon_S\right) \ .\label{U}
\end{eqnarray}
We can evaluate the difference of the fundamental energy levels $\epsilon_L-\epsilon_S$ of the liquid and solid phases by introducing the free energy $F=U-TS^{\rm BG}$. In fact, by recalling that, at equilibrium of the two phases, chemical potentials must coincide, one obtains
\begin{eqnarray}
\epsilon_L-\epsilon_S=k_B\,T_m\,\left[1+3\,\ln\left({\nu_S\over\nu_L}\right)\right]
\ ,
\end{eqnarray}
where $\mu=\partial F/\partial N$.\\
As a consequence, the melting entropy per particle $\Delta S^{\rm BG}$ is given by
\begin{equation}
\Delta S^{\rm BG}={\Lambda\over R\,T_m}=1+3\ln\left({\nu_S\over\nu_L}\right) \ ,\label{Sc}
\end{equation}
where $R=N\,k_B$ is the gas constant.\\
The melting entropy is thermally measured while frequency ratios can be obtained from electrical conductivity $k$ according to $\nu_S/\nu_L=\sqrt{k_S/k_L}$. Comparison with the observed data does not meet a wide agreement since observed ${\nu_S/\nu_L}$ are quite different from ${\nu_S/\nu_L}$ obtained using the definition above and the observed melting entropy. Better agreement can be found by means of the
semi-empirical, amended, Lindemann rule \cite{Wallace,Lawson}.\\

\section{Generalized approach}

Nonextensive statistical mechanics is a generalization of statistical mechanics for the study of anomalous systems, typically plagued by long range interactions and time persistent memory effects, whose statistical behavior may differ in a significant way from the predictions made by the orthodox theory based on the  BG entropy. In NEXT a power-law distribution is expected instead of the exponential one that is obtained by maximizing the associated nonextensive entropy. Among the different entropic forms proposed in literature \cite{CEJP,MPLB,PTP}, Tsallis entropy \cite{Tsallis} in the $(2-q)$-formalism \cite{Wada} is defined by replacing the standard logarithm in the BG entropy (\ref{SBG}) with its deformed version, the $q$-logarithm
\begin{eqnarray}
\ln_q x={x^{1-q}-1\over1-q} \ ,
\end{eqnarray}
where $q>0$ is the entropic parameter that takes into account the correlations present into the system. Consequently, the generalized entropy assumes the form
\begin{eqnarray}
S_{2-q}=-k_{\rm B}\sum_{i=1}^W{p_i^{2-q}-p_i\over1-q} \ .\label{S2q}
\end{eqnarray}
Equation (\ref{S2q}) reduces to the BG-entropy in the $q\to1$ limit: $S_1\equiv S^{\rm BG}$, as well as, in the same limit $\ln_1 x\equiv\ln x$.\\
In this way, equilibrium distribution, obtained by maximizing entropy (\ref{S2q}) under fixed mean energy and total number of particles, assumes the following form
\begin{eqnarray}
p_{j,n}={1\over\Xi_q}\exp_q\left(-\beta_q(\epsilon_j-\mu n)\right) \ ,\label{qdis}
\end{eqnarray}
where the $q$-exponential
\begin{eqnarray}
\exp_q(x)=\left[1+(1-q)\,x\right]^{1\over1-q} \ ,
\end{eqnarray}
is the inverse function of the $q$-logarithm: $\ln_q(\exp_q(x))=x$.\\
In Eq. (\ref{qdis}) $\beta_q=(\alpha\Xi_q)^{1-q}/k_{\rm B}T$, where $\alpha=(2-q)^{1/(q-1)}$ is a $q$-dependent constant and $\Xi_q$ is the $q$-deformed grand partition function.

Following NEXT approach, in the solid phase, away from the melting temperature, correlations can be neglected so that $q_S=1$. On the other hand correlations play a role in the liquid phase with $q_L\not=1$. Consequently, the crystal energy levels are those of the harmonic oscillator while the levels of the liquid phases are not rendered explicit. At melting temperature, because the melting state is a new ensemble, we expect $q_m$. The region of validity of the NEXT model is $1\leq q_m<2$.\\
Therefore, for the solid phase, we postulate the same expression of $f_S(T)$ as in the extensive theory, while for the liquid phase we make the ansatz $f_L(T)=N^{2-q}\,f_S(T)$ where correlations are accounted for by the entropic parameter.\\
At equilibrium, the non extensive entropy becomes (we refer to the Appendix for technical details)
\begin{equation}
  S_{2-q}={k_B\over2-q}\left[1+\ln_q(\alpha\Xi_q)+{U\over k_{\rm B}T}-N\ln z\right] \ ,\label{Sq}
\end{equation}
and consistently we have
\begin{eqnarray}
&&N=(\alpha\Xi_q)^{q-1}z{\partial\over\partial z}\ln\Xi_q \ ,\label{e1}\\
&&U=(\alpha\Xi_q)^{q-1}k_BT^2{\partial\over\partial T}\ln\Xi_q \ ,\label{e2}
\end{eqnarray}
while the melting heat takes the same expression (\ref{U}) obtained in the extensive case.\\
As shown in Appendix, from Eqs. (\ref{Sq})-(\ref{e2}) we obtain the relation
\begin{eqnarray}
dS_{2-q}={dU\over T}-k_{\rm B}\,dN\,\ln z \ ,
\end{eqnarray}
so that, whenever $dN=0$ it reduces to
\begin{eqnarray}
{dS_{2-q}\over dU}={1\over T} \ ,\label{Sqq1}
\end{eqnarray}
at the base of the Legendre structure of the theory \cite{Wada1}.
Furthermore, under the {\em no work} condition, we obviously must have $\Delta U=\delta Q$ so that Eq. (\ref{Sqq1}) reduces to the fundamental relation
\begin{eqnarray}
dS_{2-q}={\delta Q\over T} \ ,
\end{eqnarray}
satisfied by the thermodynamical entropy.\\
Again, at melting the two chemical potentials $\mu_S$ and $\mu_L$, with $\mu=\partial F_{2-q}/\partial N$, must be equal, where $F_{2-q}=U-T\,S_{2-q}$. From these definitions one obtains the expression of the nonextensive melting entropy
\begin{eqnarray}
  \Delta S_{2-q}={1\over2-q}\left(\Delta S^{\rm BG}+C_{2-q}(N)\right) \ .\label{sQq}
\end{eqnarray}
The corrective term $C_{2-q}(N)$ reads
\begin{eqnarray}
C_{2-q}(N)=\ln\left({2-q\over N^{q-1}}\right)+{w^2\over1+w} \ ,\label{final}
\end{eqnarray}
with $w=W((q-1)N/(2-q))$, where $W(x)$ is the Lambert function defined as $W(x)e^{W(x)}=x$. We remark that the nonextensive entropy melting per particles $\Delta S_{2-q}$ depends, through the term $C_{2-q}(N)$, on the number of particles $N$ of the system. We discuss the importance of this fact in the next paragraph.\\
In the $q\to1$ limit the corrective term vanishes for all $N$, $C_1(N)=0$, so that $\Delta S_{2-q}\to\Delta S^{\rm BG}$.

\section{Results and discussion}

We can derive the value of $q_m$ in order to obtain the agreement between the experimental value of melting entropy and $\Delta S_{2-q}$ of Eq. (\ref{sQq}) evaluated in terms of the observed values of the quantities $\nu_S,\,\nu_L$ and $T_m$.\\
\begin{table}[t]
\center{Normal metals $(N=10^{20})$.}
\item[]\begin{tabular}{@{}cccccc}
\hline
 \ $Z$ \ & \ element \ &  \ \ $\Delta S$ \ & \ $\nu_S/\nu_L$ \ & \ $T_m(K)$ \ & \ $q_m$\\
\hline
3& Li & 0.80 & 1.30 & 453.7 & 1.900\\
11 & Na & 0.85 & 1.20 & 371.0 & 1.893\\
13 & Al & 1.39 & 1.28 & 933.0 & 1.875\\
19 & K & 0.83 & 1.24 & 336.4 & 1.896\\
29 & Cu & 1.16 & 1.43 & 1357 & 1.891\\
30 & Zn & 1.28 & 1.45 & 692.7 & 1.887\\
37 & Rb & 0.84 & 1.27 & 312.6 & 1.897\\
47 & Ag & 1.11 & 1.38 & 1234 & 1.891\\
55 & Cs & 0.83 & 1.29 & 301.6 & 1.898\\
79 & Au & 1.14 & 1.54 & 1336 & 1.897\\
80 & Hg & 1.18 & 1.79 & 234.3 & 1.905\\
81 & Tl & 0.86 & 1.41 & 577 & 1.903\\
82 & Pb & 0.96 & 1.44 & 600.6 & 1.900\\
\hline
\end{tabular}
\caption{For 13 metals the observed values of melting entropy, crystal over liquid frequency and melting temperature are reported with the values of $q_m$ deduced from Eq. (\ref{sQq}) for $N=10^{20}$ mean particle numbers.}
\end{table}
\begin{table}[t]
\center{Anomalous metals $(N=10^{20})$.}
\item[]\begin{tabular}{@{}cccccc}
\hline
 \ $Z$ \ & \ element \ &  \ \ $\Delta S$ \ & \ $\nu_S/\nu_L$ \ & \ $T_m(K)$ \ & \ $q_m$\\
\hline
14 & Si & 3.61 & 1.31 & 1685 & 1.791\\
31 & Ga & 2.23 & 0.76 & 302.9 & 1.808\\
32 & Ge & 3.68 & 1.21 & 1210 & 1.783\\
51 & Sb & 2.64 & 1.22 & 904 & 1.823\\
83 & Bi & 2.50 & 1.16 & 544.5 & 1.825\\
\hline
\end{tabular}
\caption{For 5 non-metals the observed values of melting entropy, crystal over liquid frequency and melting temperature are reported with the values of $q_m$ deduced from Eq. (\ref{sQq}) for $N=10^{20}$ mean particle numbers.}
\end{table}
In Tables I and II are reported the values of $q_m$ for several elements needed to obtain the observed values of $\Delta S$, in units of $k_B$ per atoms, by means of the non-extensive expression (\ref{sQq}), for $N=10^{20}$ (the samples used in the experiments contain approximatively this number of atoms). All values of $\Delta S$, $\nu_S/\nu_L$ and $T_m$ are the measured values collected by Wallace \cite{Wallace} from many different experimental works quoted in his papers, and by others \cite{Regel,Chi}. We recall that Si and Ge are semiconductors when solid, and metals when liquid. Sb and Bi are semimetals in the solid state and metals in the liquid state. Finally, Ga forms metal when solid, becomes metal when liquid but with anomalous and  almost unknown change in the electric ground state from solid to liquid.\\
Experimental results on $\Delta S$ (or $\Lambda/R\,T_m$) and $\nu_S/\nu_L$ come from thermal and electric conductivity measurements, respectively \cite{Wallace,Fowler2,Hultgren,Chi}.\\
In the normal metals, melting entropy is composed by a contribution of the lattice dynamics and one due mainly to liquid correlations. In the anomalous metals (Si, Ga, Ge, Sb, Bi) melting entropy is much greater than normal: change in the electronic energy is due to a change of electronic ground state upon melting.\\
Wallace \cite{Wallace} has given an expression of metal melting entropy which reads
\newpage
\begin{eqnarray}
\nonumber
  \Delta S&=&\left({\Lambda\over
  R\,T_m}\right)=\Big(S_O^L+S_C^L+S_Q^L+S_E^L\Big)\\
  & &-\Big(S_H^S+S_A^S+S_Q^S+S_E^S\Big) \ ,\label{ds}
\end{eqnarray}
where $S^L$ and $S^S$ indicate the liquid and solid entropy, respectively, index $O$ means one-body contribution, $C$ correlation contribute, $H$ the solid harmonic term, $A$ the anharmonic,
$Q$ quantum and $E$ electronic contribution, obtaining agreement within experimental errors.\\
\begin{figure}[t]
\resizebox{140mm}{!}{\includegraphics{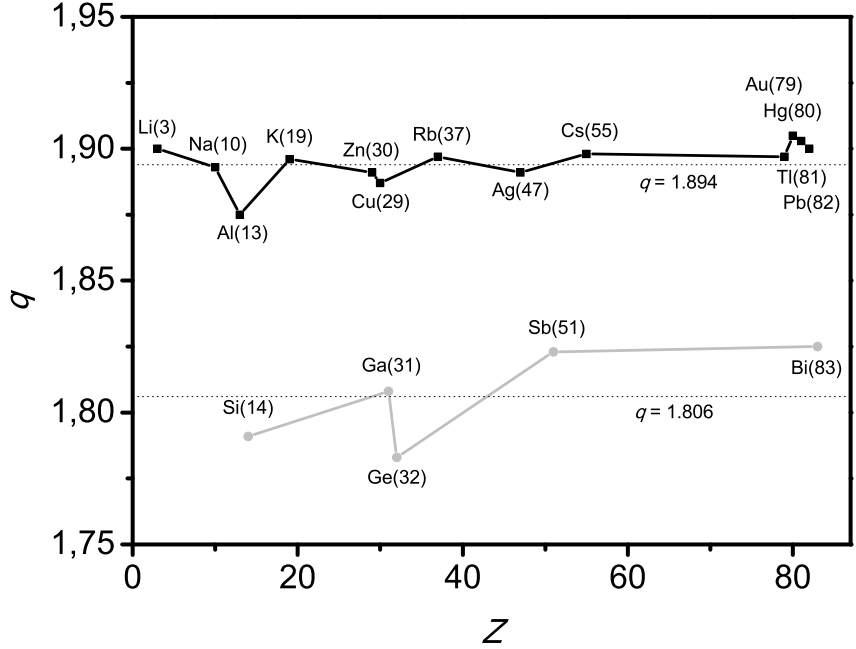}}
  \caption{The entropic parameter $q$ versus $Z$ of the elements from Li to Bi. Metals and non-metals are clearly separated.}
\end{figure}
\begin{figure}[t]
\resizebox{140mm}{!}{\includegraphics{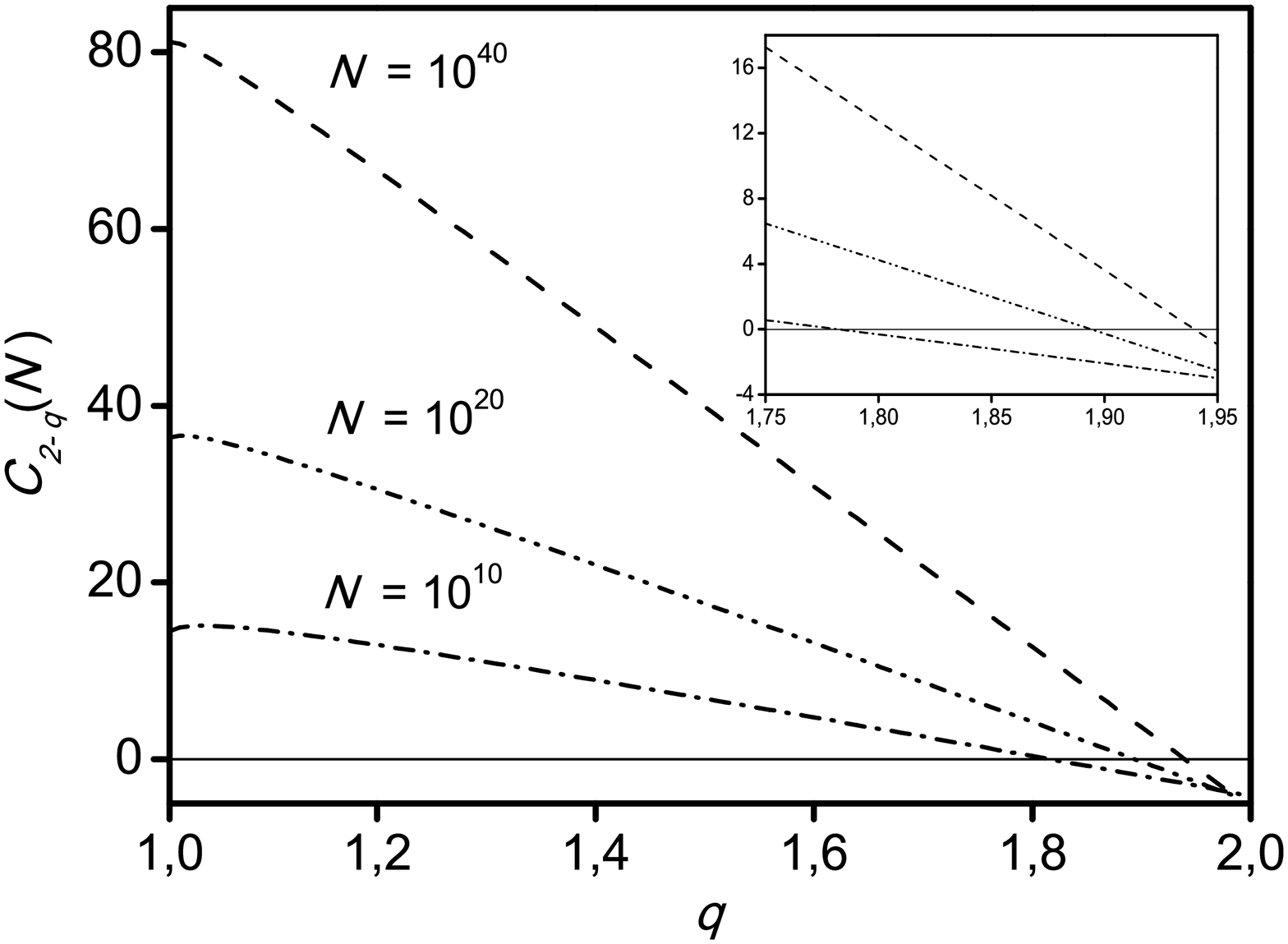}}
  \caption{Shape of $C_{2-q}(N)$ vs $q$ for several value of mean particle numbers $N$. The inset reproduces the magnification of the region of $q$ corresponding to the Tables I and II.}
\end{figure}
We may see that $S_O^L-S_H^S$ can be related to the classical melting entropy $\ln e\,(\nu_S/\nu_L)^3$.
In the normal metals the contribution from $Q$ and $E$ terms is nearly negligible, while the $S^L_C$ and $S^S_A$ quantities are very important and the value of $q_m$ strictly depends on them.\\
The use of NEXT statistical mechanics avoids the calculation of correlation and other different contributions. The values of $q_m$ we obtain setting Eq. (\ref{sQq}) equal to the observed $\Delta S$ are those reported in Table I and II and clearly depicted in Fig. 1, where we report for 13 normal metals and 5 anomalous metals the value of $q_m$ given in Table I and II.\\
From experimental data we see that metals can be separated in two classes respect the parameter $q_m$. All normal metals show a value of the entropic parameter $q_m$ between $1.88$ and $1.91$ (mean value $1.894$) while anomalous metals between $1.79$ and $1.83$ (mean value $1.806$). Although the difference of the values of $q_m$ for normal metals and anomalous metals is only of few percent it is decisive enough to recover the observed normal metal melting entropies (average value $1.15$) that sensibly differ from the observed anomalous metal melting entropies (average value $3.11$). In fact, as it happens in other situations, very small differences of $q$ can produce large differences in physical observables (for instance, in the rates of thermonuclear reactions).\\ However, in our opinion, the discrepancy in the value of $q_m$ between normal and anomalous metals arise from the different correlations present inside the two systems.\\
The sensitivity of $\Delta S_{2-q}$ from the $q$ parameter is better clarified in Fig. 2 where we show the dependence of $C_{2-q}(N)$ from $q$ for several values of $N$. As $N$ increases the sensibility of $C_{2-q}(N)$ reduces. For instance, at $q_m=1.5$ the variation of $C_{2-q}(N)$ in the range of $N\approx10^{20}\div10^{40}$ (twenty orders!) is about $150\%$ while for $N\approx10^5\div10^8$ (increasing of only three orders of magnitude) the variation of $C_{2-q}(N)$ is about $120\%$. Thus, in the thermodynamical limit of large $N$ melting entropy for particle is almost independent from $N$ and we expect that entropy scales as $\Delta S(N)\approx N\Delta S(1)$. Therefore, in this limit, entropy per particles (\ref{sQq}) does not depend by $N$, i.e. melting entropy is an extensive quantity only for $N\to\infty$. We remark that, although $C_{2-q}(N)$ is a continuous function, its slope, for $q\to1$, is almost vertical, so that $C_{2-q}(N)\to0$ in the same limit.\\
\begin{figure}[t]
\resizebox{140mm}{!}{\includegraphics{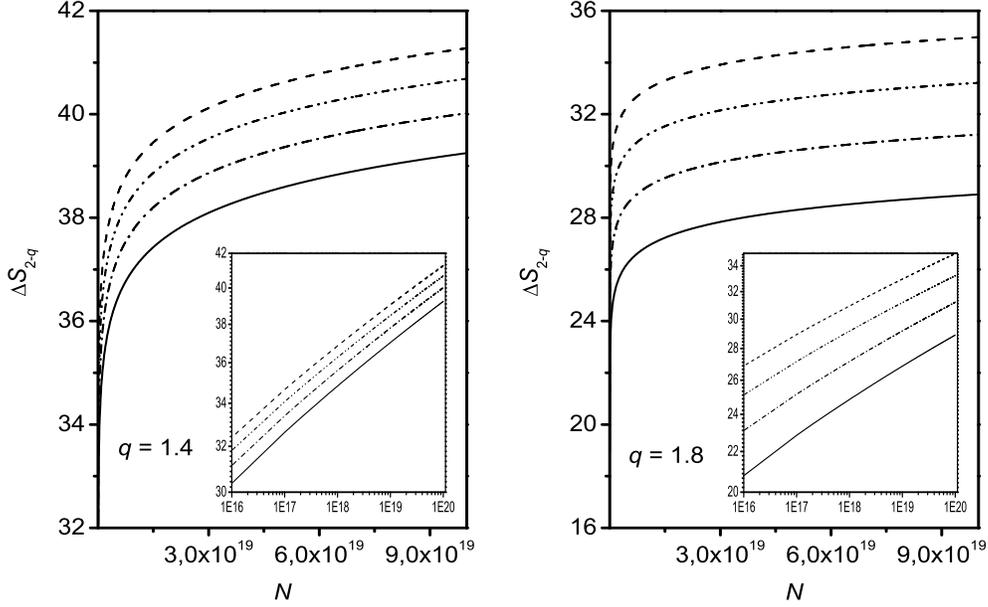}}
  \caption{Plot of $\Delta S_{2-q}$ vs $N$ for several values of the crystal over liquid frequency (solid line $\nu_S/\nu_L=1.2$, dash-dot line $\nu_S/\nu_L=1.4$, dash-dot-dot line $\nu_S/\nu_L=1.6$ and dash line $\nu_S/\nu_L=1.8$) and $q$. In the insets the same figures in the log-log scale.}
\end{figure}
In Fig. 3, $\Delta S_{2-q}$, for four crystal over liquid frequencies, is plotted versus $N$ for two values of $q_m$. In the log-log scale of the two insets we can see that NEXT melting entropies $\Delta S_{2-q}$ behave, in first approximation, like parallel straight lines, indicating that $\Delta S_{2-q}$ can also be expressed as a power law function of $N$ according to $\Delta S_{2-q}\approx \alpha\,N^\beta$ with $\beta\to0$ for large $N$.

In Fig. 4, $\Delta S_{2-q}$ is plotted versus $q$ for few values of $\nu_S/\nu_L$ and for several values of $N$. It is shown that the difference among the curves, in the region between $q_m=1.79$ and $q_m=1.91$, decreases as $N$ increases. For $q=1$ we have $\Delta S_{2-q}=\Delta S^{\rm BG}$.
\begin{figure}[t]
\resizebox{140mm}{!}{\includegraphics{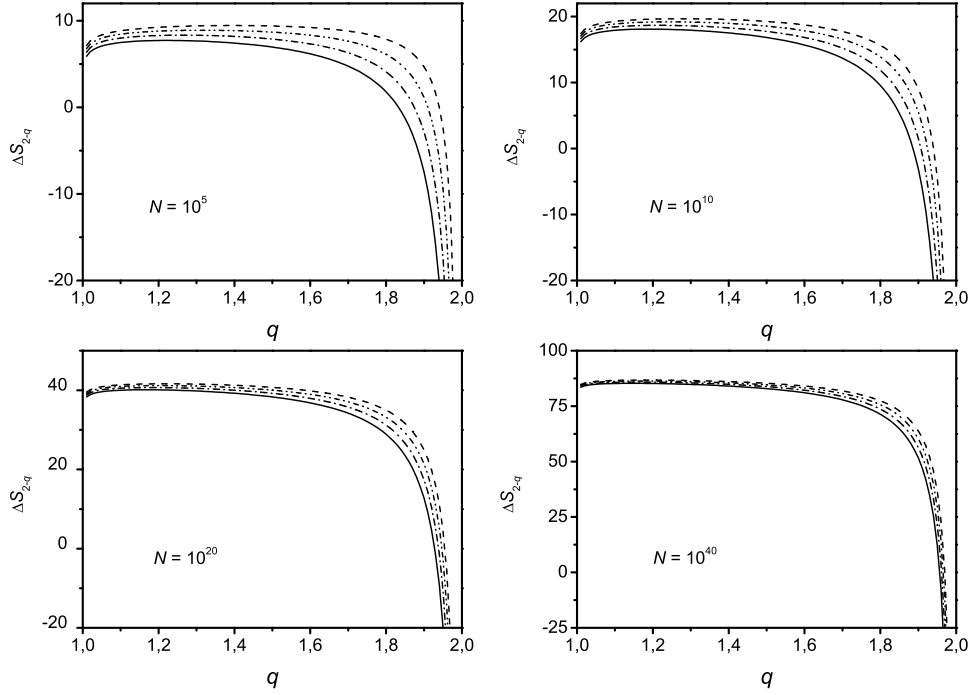}}
  \caption{Plot of $\Delta S_{2-q}$ vs $q$ for several values of the crystal over liquid frequency (solid line $\nu_S/\nu_L=1.2$, dash-dot line $\nu_S/\nu_L=1.4$, dash-dot-dot line $\nu_S/\nu_L=1.6$ and dash line $\nu_S/\nu_L=1.8$) and for four different values of $N$.}
\end{figure}
From Fig. 4 we can see that once $N$ and $\nu_S/\nu_L$ are fixed, a value of $q^\ast_m$ smaller than 2 exists such that, for $q_m>q^\ast_m$, the melting entropy $\Delta S_{2-q}$ becomes negative. Therefore, we limit ourselves to consider the NEXT metal melting model valid in the range $1\leq q<q^\ast_m$.\\ Finally, let us remark that, because the gap between the two average values of $q$ (for metals and non-metals) increases as the number of particles $N$ diminishes, the approach described in this work is particularly suitable for the melting of nanosystems.

\section{Conclusions}
In this work, we have reconsidered the Mott relation of melting entropy. We have modified and generalized it by taking into account that in metal melting the system transfers from a localized, metastable state with no correlation to a non-localized, highly correlated liquid state. Therefore, the use of generalized NEXT statistics is justified and appropriate for a precise description of the metal melting entropy. The division of the metals into two groups, as already indicated in the past by Regel and Glazov and  by Wallace, with two different values of the parameter $q$ has been recovered, indicating that $q$ could be used as a discriminator parameter between normal and anomalous metals.
Of course, the values of the experimental quantity $\Delta S$ discriminate metals from nonmetals too. Hoverer, $\Delta S$ and $q$ are two distinct quantities (the former experimental and the latter theoretical) with distinct roles. By knowing ``a priori'' the values of $q$ for metals and for nonmetals, one could give the appropriate value of $q$ to any element before measuring $\Delta S$. We have found $q$ ``a posteriori'' from experimental data of several elements so that any other element belonging to one of the two categories must have a $q$ value belonging to the corresponding class. In this way, the theoretical value of melting entropy arising from Eq. (19) gives us $\Delta S$.\\
The physical meaning of the entropic parameter $q$ can be understood from the microscopic description of the Wallace works \cite{Wallace} in terms of the two-body correlation function. Depending on the two values of $q$ other metals, whose melting entropy is not reported here, belong to one of the two classes.\\  Melting of metals is still today widely studied to understand properties of melting in condensed matter, semiconductors, superconductors, nanostructure and also for its importance in planetary and stellar physics. Calculations of melting (and freezing) entropy of light elements  (H, He) and medium-heavy (Fe, Si, Mg) elements in the Sun, Jupiter and Saturn are useful for the knowledge of the dynamics of their inner cores \cite{Nellis}.
Furthermore, concerning the inner core of the Earth \cite{Gubbins,Luo,Chau}, it has been recently observed that in the sub-mantle there is simultaneously melting and freezing because of circulation of heat in the overlying rocky mantle with implications in the terrestrial magnetic field and seismology.\\
The study of melting entropy in nanostructures, semiconductors and superconductors, in neutron stars, white and brown dwarfs and quark-gluon phase \cite{Lavagno1,Lavagno2,Corraddu} of compact stars can be a field of application of the expression of melting (and freezing) entropy of metals derived in this work by means of generalized NEXT statistical mechanics. Therefore new highly precise measurements of the metal melting entropy are greatly welcome.

\app
\section{}

In this appendix we give a sketch for the derivation of Eq. (\ref{sQq}) remanding to the relevant literature for the details.\\
The starting point is the $q$-entropy (\ref{S2q}) here rewritten in
\begin{eqnarray}
S_{2-q}=-k_{\rm B}\sum_{i,m}p_{i,m}\ln_qp_{i,m} \ .\label{Sq1}
\end{eqnarray}
Equilibrium distribution can be obtained from the following variational problem
\begin{eqnarray}
{\delta\over\delta p_{j,n}}\left[S_{2-q}-\sum_{i,m}\big(\gamma-\beta\epsilon_i-\lambda m\big)p_{i,m}\right]=0 \ ,
\end{eqnarray}
where $\gamma$, $\beta$ and $\lambda$ are the Lagrange multipliers related to the normalization, the mean energy and the mean particle numbers, respectively.\\
Ultimately, the distribution takes the form
\begin{eqnarray}
\nonumber
p_{j,n}&=&\alpha\left[1-{1-q\over k_{\rm B}}\Big(\gamma+\beta\epsilon_j+\lambda n\Big)\right]^{1\over1-q}\\
&=&{1\over\Xi_q}\exp_q\left(-\beta_q\Big(\epsilon_j-\mu n\Big)\right) \ ,\label{dist}
\end{eqnarray}
where $\exp_q(x)=[1+(1-q)x]^{1/(1-q)}$ is the $q$-exponential,
\begin{eqnarray}
\Xi_q={1\over\alpha}\left(1-{1-q\over k_{\rm B}}\,\gamma\right)^{1\over q-1} \ ,\label{grand}
\end{eqnarray}
is the grand partition function, $\alpha=(2-q)^{1/(q-1)}$ is a constant, $\beta=1/T$, $\beta_q=(\alpha\Xi_q)^{1-q}/k_{\rm B}T$ and $\mu=-\lambda/\beta$.
Furthermore, from Eq. (\ref{grand}) we obtain
\begin{eqnarray}
\gamma=k_{\rm B}\ln_{2-q}\left(\alpha\Xi_q\right) \ ,
\end{eqnarray}
where $\ln_{2-q}(x)=(x^{q-1}-1)/(q-1)$.\\
By inserting Eq. (\ref{dist}) in (\ref{Sq1}) we can rewrite the $q$-entropy as
\begin{eqnarray}
S_{2-q}={k_{\rm B}\over2-q}\Bigg[1+\ln_{2-q}\left(\alpha\Xi_q\right)+{U\over k_{\rm B}T}-N\ln z\Bigg] \ , \label{questa}
\end{eqnarray}
where $z=e^{\mu/k_{\rm B}T}$ is the fugacity.\\
To evaluate the quantity $dS_{2-q}$ we can proceed in two different ways:\\ 1) by differentiating directly Eq. (\ref{questa})
\begin{eqnarray}
\nonumber
dS_{2-q}&=&{k_{\rm B}\over2-q}\,\left[(\alpha\Xi_q)^{q-1}d\ln\Xi_q+d\left({1\over k_{\rm B}T}\right)U
+{dU\over k_{\rm B}T}\right.\\
&-&\Bigg.dN\ln z-Nd\ln z\Bigg] \ , \label{dS1}
\end{eqnarray}
2) by accounting for the relation $d(x\ln_q x)=\ln_q(x/\alpha)dx$, from Eq. (\ref{Sq1}), under the condition of no work
\begin{eqnarray}
dS_{2-q}={dU\over T}-k_{\rm B}dN\ln z \ .\label{dS2}
\end{eqnarray}
By matching Eq. (\ref{dS1}) with Eq. (\ref{dS2}) we can derive the following
relations
\begin{eqnarray}
&&U=k_{\rm B}(\alpha\Xi_q)^{q-1}T^2{\partial\over\partial T}\ln\Xi_q \ ,\label{Uq}\\
&&N=(\alpha\Xi_q)^{q-1}z{\partial\over \partial z}\ln\Xi_q \ ,\label{Nq}
\end{eqnarray}
together with the well known relation $dU/dN=\mu$. They state, consistently, the Legendre structure of the theory.\\
In order to obtain the explicit expression for $U_S$ and $U_L$ we employ the grand partition function
for the solid and liquid phase, given by $\Xi_S=(1-zf_S)^{-1}$ and $\Xi_L=\exp(zf_L)$, respectively, (hereinafter we omit the index $q$ for sake of exposition), where $f_S=\left(k_BT/h\nu_S\right)^3\exp\left(-\epsilon_S/k_BT\right)$
and $f_L=N^{2-q}\left(k_BT/h\nu_L\right)^3\exp\left(-\epsilon_L/k_BT\right)$.
We have
\begin{eqnarray}
&&U_S=\left(3k_{\rm B}T+\epsilon_S\right)N_S \ ,\label{Us}\\
&&U_L=\left(3k_{\rm B}T+\epsilon_L\right)N_L \ ,\label{Ul}
\end{eqnarray}
where
\begin{eqnarray}
&&N_S=(2-q){zf_S\over(1-zf_S)^q} \ ,\label{ns}\\
&&N_L=(2-q)zf_Le^{(q-1)zf_L} \ ,\label{nl}
\end{eqnarray}
obtained from  Eq. (\ref{Nq}). In the following we pose $N_S=N_L=N$.\\ Let us to observe that Eqs. (\ref{Us}) and (\ref{Ul}) formally coincide with the corresponding relations derived in the standard BG-theory.\\
The final step follows by introducing the free energy
\begin{eqnarray}
\nonumber
&&F_{2-q}=U-TS_{2-q}\\
&&=-{k_{\rm B}T\over2-q}\Bigg[1+\ln_{2-q}\left(\alpha\Xi_q\right)-(1-q){U\over k_{\rm B}T}-N\ln z\Bigg] \ ,
\end{eqnarray}
with the related relation
\begin{eqnarray}
\mu={\partial F_{2-q}\over\partial N} \ .
\end{eqnarray}
After a bit of algebras we obtain
\begin{eqnarray}
&&\mu_S={k_{\rm B}T\over2-q}\left[3\ln{h\nu_S\over k_{\rm B}T}+{\epsilon_S\over k_{\rm B}T}+(1-q)\left(3+{\epsilon_S\over k_{\rm B}T}\right)\right] \ ,\label{mus}\\
\nonumber
&&\mu_L={k_{\rm B}T\over2-q}\left[3\ln{h\nu_L\over k_{\rm B}T}+{\epsilon_L\over k_{\rm B}T}+(1-q)\left(3+{\epsilon_L\over k_{\rm B}T}\right)\right.\\
&&\hspace{10mm}\left.-\ln{2-q\over N^{q-1}}-{1\over1+w}-w\right] \ ,
\end{eqnarray}
where, we posed $w=(q-1)zf_L$ and, reasonably, we assumed $zf_S\simeq1$, as it follows from (\ref{ns}).\\
The value of $w$ is given implicitly from Eq. (\ref{nl}), that rewritten in
\begin{eqnarray}
  {q-1\over2-q}N=we^w \ ,
\end{eqnarray}
it becomes a transcendant equation whose solution $w\equiv W((q-1)N/(2-q))$ is known in literature as Lambert function.\\
At equilibrium, $\mu_L=\mu_S$ and we obtain
\begin{eqnarray}
\epsilon_L-\epsilon_S={k_{\rm B}T\over2-q}\left(3\ln{\nu_S\over\nu_L}+\ln{2-q\over N^{q-1}}+{1\over1+w}+w\right) \ . \label{ultima}
\end{eqnarray}
Taking into account Eqs. (\ref{Us}), (\ref{Ul}) and (\ref{ultima}) the nonextensive melting entropy per particle is then given by
\begin{eqnarray}
\nonumber
\Delta S_{2-q}&=&{U_L-U_S\over Nk_{\rm B}T}\\
&=&{1\over 2-q}\left(1+3\ln{\nu_S\over\nu_L}+\ln{2-q\over N^{q-1}}+{w^2\over1+w}\right) \ ,
\end{eqnarray}
that coincides with Eq. (\ref{sQq}) given in the text.


\end{document}